\documentclass{article}

\usepackage{packages/arxiv}
\usepackage{packages/main}

\usepackage[inline]{packages/trackchanges} 
\addeditor{Davide}

\graphicspath{{figures/}}

\makeatletter
\def\input@path{{src/}}
\makeatother

\title{On the range of validity of parabolic models for fluid flow through isotropic homogeneous porous media}

\author{
    \href{https://orcid.org/0000-0002-3442-6857}{\includegraphics[scale=0.06]{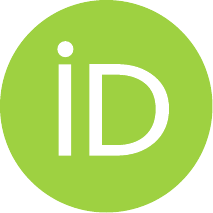}\hspace{1mm}Davide Dapelo}\thanks{Corresponding author: \texttt{d.dapelo@liverpool.ac.uk}}\\
	Department of Civil and Environmental Engineering,\\
    University of Liverpool,\\
    Liverpool, UK \\
	\And
    \href{https://orcid.org/0000-0001-8555-4245}{\includegraphics[scale=0.06]{orcid.pdf}\hspace{1mm}Stephan Simonis}\\
	Institute for Applied and Numerical Mathematics,\\
    Karlsruhe Institute of Technology,\\
    76131 Karlsruhe, Germany \\
	\And
    \href{https://orcid.org/0000-0002-0625-439X}{\includegraphics[scale=0.06]{orcid.pdf}\hspace{1mm}Mohaddeseh Mousavi Nezhad}\\
    Department of Civil and Environmental Engineering, \\
    University of Liverpool, \\
    Liverpool, UK \\
	\And
    \href{https://orcid.org/0000-0003-1026-6462}{\includegraphics[scale=0.06]{orcid.pdf}\hspace{1mm}Mathias J. Krause}\\
    Lattice Boltzmann Research Group, \\
    Karlsruhe Institute of Technology, \\
    76131 Karlsruhe, Germany \\
	\And
    \href{https://orcid.org/0000-0001-8348-5004}{\includegraphics[scale=0.06]{orcid.pdf}\hspace{1mm}John Bridgeman}\\
    Department of Civil and Environmental Engineering, \\
    University of Liverpool, \\
    Liverpool, UK \\
}

\renewcommand{\headeright}{Dapelo et al.}
\renewcommand{\shorttitle}{Range of validity of parabolic REV}

\hypersetup{
pdftitle={REV},
pdfsubject={},
pdfauthor={Davide Dapelo}
}

\begin{document}
\maketitle

\begin{abstract}
    Lattice-Boltzmann methods are established mesoscopic numerical schemes for fluid flow, that recover the evolution of macroscopic quantities (\viz velocity and pressure fields) evolving under macroscopic target equations. The approximated target equations for fluid flows are typically parabolic and include a (weak) compressibility term. A number of Lattice-Boltzmann models targeting, or making use of, flow through porous media in the representative elementary volume, have been successfully developed. However, apart from two exceptions, the target equations are not reported, or the assumptions for and approximations of these equations are not fully clarified.

    Within this work, the underlying assumption underpinning parabolic equations for porous flow in the representative elementary volume, are discussed, clarified and listed. It is shown that the commonly-adopted assumption of negligible hydraulic dispersion is not justifiable by clear argument---and in fact, that by not adopting it, one can provide a qualitative and quantitative expression for the effective viscosity in the Brinkman correction of Darcy's law. Finally, it is shown that, under certain conditions, it is possible to interpret porous models as Euler-Euler multiphase models wherein one phase is the solid matrix.
\end{abstract}
\keywords{
Representative Elementary Volume
\and
Navier-Stokes
\and
Volume-Averaging
\and
Porous Media
}

\subjclass{76S05, 35Q30}

\begin{mdframed}
\section*{Nomenclature}
\nopagebreak
\small
\begin{longtable}{lp{0.85\textwidth}}
CFD & Computational Fluid Dynamics \\
EE & Euler--Euler \\
LES & Large Eddy Simulations \\
RANS & Reynolds-Averaged Navier--Stokes \\
REV & Representative Elementary Volume \\

$\mat{\cdot\,}$ & $\cdot$ referred to the microscopic matrix \\
$\rev{\cdot\,}$ & $\cdot$ referred to the representative elementary volume \\
$\sys{\cdot\,}$ & $\cdot$ referred to the macroscopic system \\
$\avg{\cdot}$ & Surface phase average \\
$\avgf{\cdot}$ & Intrinsic phase average \\
$\dv{\cdot}$ & Spatial deviation \\
${\cdot\,}^p$ & (in the EE context) $\cdot$ referred to the phase $p$ \\

$\Delta$ & Smagorinsky filter width, \si{\meter} \\
$\Phi_i^{+/-}\left[\Lambda\right]$ & Flux of $\Lambda$ across the $i$ direction, positive or negative versus, \si{\left\{\Lambda\right\}\per\second} \\

$\alpha^p$ & (in the EE context) Phase fraction of the phase $p$ \\
$\varepsilon$ & Porosity \\
$\mu$ & Dynamic viscosity, \si{\pascal\second} \\
$\mu_e$ & Effective dynamic viscosity, \si{\pascal\second} \\
$\rho$ & Liquid phase density \si{\cubic\meter\per\second} \\
$\rhomat$ & Microscopic liquid phase density \si{\cubic\meter\per\second} \\
$\tau$ & Shear stress, \si{\pascal} \\
$\chi^p$ & (in the EE context) Indicator function of the phase $p$ \\

$\vect{B}$ & Body force, \si{\newton} \\
$C_s$ & Smagorinsky constant \\
$F_\varepsilon$ & Geometric function \\
$\tens{I}$ & Identity tensor \\
$K$ & Porous permeability (scalar), \si{\square\meter} \\
$\tens{K}$ & Porous permeability (tensor), \si{\square\meter} \\
$\Kn$ & Knudsen number \\
$M$ & Mass, \si{\kilogram} \\
$\vect{M}^{ab}$ & (in the EE context) Momentum transfer from phase $a$ to phase $b$, \si{\newton\per\cubic\meter} \\
$\vect{P}$ & Momentum, \si{\kilogram\meter\per\second} \\
$\tens{R}$ & Reynolds-stress-like tensor, \si{\pascal} \\
$\Remat$ & Microscopic Reynolds number \\
$\vect{S}$ & Surface force, \si{\newton} \\
$\tens{S}$ & Shear rate, \si{1/\second} \\
$V$ & (in the EE context) Control volume for the volume-averaging procedure, \si{\cubic\meter} \\
$V^p$ & (in the EE context) Portion of the control volume occupied by the phase $p$, \si{\cubic\meter} \\
$\Vrev$ & Representative elementary volume, \si{\cubic\meter} \\
$\Vrevf$ & Portion of the representative elementary volume occupied by the fluid, \si{\cubic\meter} \\
$\partial\Vrevf$ & Interfacial area between fluid phase and porous matrix within $\Vrev$, \si{\square\meter} \\

$a$ & Flow fraction \\
$\vect{b}$ & Body force density, \si{\newton\per\cubic\meter} \\
$\bmat$ & Microscopic body force density, \si{\newton\per\cubic\meter} \\
$\ivers$ & Versor along the $x$ direction \\
$\jvers$ & Versor along the $y$ direction \\
$\kvers$ & Versor along the $z$ direction \\
$k_1$ & Inertial permeability, \si{meter} \\
$\lmat$ & Length scale of the microscopic porous matrix, \si{\meter} \\
$\lrev$ & Length scale of the representative elementary volume, \si{\meter} \\
$\lsys$ & Length scale of the macroscopic system, \si{\meter} \\
$\dot{m}^{ab}$ & (in the EE context) Mass transfer rate from phase $a$ to phase $b$, \si{\kilogram\per\cubic\meter\per\second}\\
$\nvers$ & Generic versor \\
$p$ & Pressure, \si{\pascal} \\
$\pmat$ & Microscopic pressure, \si{\pascal} \\
$\usurf$ & Darcy's flux, or surface velocity, \si{\meter\per\second} \\
$t$ & Time, \si{\second} \\
$\tmat$ & Microscopic time scale, \si{\second} \\
$\tsys$ & Macroscopic time scale, \si{\second} \\
$\useep$ & Averaged liquid phase seepage velocity, \si{\meter\per\second} \\
$\umat$ & Liquid phase microscopic velocity, \si{\meter\per\second} \\
$\vect{x}$ & Spatial position, \si{\meter} \\
\end{longtable}
\normalsize
\end{mdframed}
\tableofcontents
\makeatletter
\g@addto@macro\input@path{{src/introduction/}}
\makeatother

\section{Introduction}
\label{s:intro}
The flow of fluid through an isotropic porous medium was first investigated by \cite{Darcy1856}, who found an empirical linear law to relate the liquid phase flux across a unit surface cutting both liquid phase and solid matrix (``Darcy's flux'', or surface velocity) with the gradient of the liquid phase pressure:
\begin{equation}\label{eq:darcy}
    \usurf = -\frac{K}{\mu}\nabla p . 
\end{equation}
The Darcy flux can be related to the seepage velocity $\useep$ of the liquid phase, averaged across a volume known as Representative Elementary Volume (REV) that is much larger than the porous matrix, and at the same time much smaller than the physical extension of the smallest variation of the flow fields of interest for the problem:
\begin{equation}\label{eq:darcyFlux}
    \usurf = \varepsilon\useep . 
\end{equation}
This law is found to be valid for homogeneous flows with microscopic Reynolds numbers (\viz defined on the pore length) $\mat{\Re}$ up to 10 \citep{Bear1998}. A correction quadratic on the Darcy flux was proposed by \cite{Forchheimer1901} for $10<\mat{\Re}<100$:
\begin{equation}\label{eq:forchheimer}
    \nabla p = -\frac{\mu}{K}\usurf - \frac{\rho}{k_1}\left|\usurf\right|\usurf
\end{equation}
and an extension containing a dissipative term proportional to the liquid phase's effective dynamic viscosity was proposed by \cite{Brinkman1947} to reconcile the model with the Stokes equation for small values of permeability:
\begin{equation}\label{eq:brinkman}
    \mu_e\nabla^2\usurf + \usurf = -\frac{K}{\mu}\nabla p . 
\end{equation}

This set of models has found a wide range of applicability, especially in the fields of hydrogeology \citep{Bear1998,Hiscock2006} and petroleum engineering \citep{Ahmed2010}. A number of theoretical derivations have also been proposed---the range of the theoretical techniques employed includes volume averaging \citep{whitakerMethodVolumeAveraging1999} and asymptotic analysis \citep{Giorgi1997}. These results been limited to the stationary incompressible case, partly because of the stationary nature of the above-mentioned applications. A partial exception is represented by \cite{Bear1990}, but only a conservation equation for the mass was presented for the general case of time-dependent compressible flow. Further work on volume averaging analyzed incompressible, time-varying flow---specifically, scalar advection \citep{Wood2013} and velocity advection in a linearized problem \citep{Lasseux2019}---and demonstrated that unsteady processes in a hierarchical system induce a memory effect that can only be captured by means of a convolution product between an apparent permeability tensor and the forcing term.

\subsection{Parabolic equations and Lattice-Boltzmann modelling}
\label{s:parabolic}
In relatively recent times, a number of numerical models for porous flow have been developed within the Lattice-Boltzmann numerical method \citep{Krueger2016} to simulate flow in porous media \citep{Spaid1997,Guo2002,Wang2015}, and also as a part of the Homogenised Lattice-Boltzmann Method (HLBM), a method to simulate immersed resolved objects \citep{Krause2017,Trunk2018,Dapelo2019,Trunk2021,Marquardt2023,Simonis2025}. However, the above-mentioned literature reports simulations of transient flows, which fall outside the remit of the stationary models proposed above (Equations~\ref{eq:darcy}, \ref{eq:forchheimer} and~\ref{eq:brinkman}). This is due to the fact that the Lattice-Boltzmann is a mesoscopic method---meaning that the method resolves a continuity (Lattice-Boltzmann) equation for a mesoscopic quantity (the one-particle probability density function in the phase space); the observable fields are recovered as moments of the probability density function, and are shown to obey a ``target'' macroscopic equation through a multi-scale (``Chapman-Enskog'') expansion \citep{Krueger2016}. The Lattice-Boltzmann model for a given problem is either built \emph{bottom-up} by defining a Lattice-Boltzmann equation and showing through an asymptotic expansion \citep{Krueger2016} that this is equivalent to solving the desired target; or derived \emph{top-down} by working out the Lattice-Boltzmann equation from a desired target equation \citep{Simonis2020,Simonis2023,Simonis2022,Simonis2023pde}. In both cases, the target equation is a parabolic partial differential equation that is first-order in time and second-order in space. As the conserved moments (density and velocity) of the native Lattice-Boltzmann equation for fixed numerical parameter configurations are approximated solutions of a weakly compressible fluid flow, the time-varying models mentioned above \citep{Wood2013,Lasseux2019} are not immediately applicable to the purpose of supporting Lattice-Boltzmann models because of the non-local nature of the convolutional terms and the assumed incompressibility and linearity.

In \cite{Guo2002}, the following macroscopic equation is targeted, after explicitly ignoring the compressibility terms:
\begin{equation}\label{eq:guo-incompressible}
\begin{aligned}
    \nabla\cdot\usurf &= 0 , \\
    \rho\partial_t\usurf + \rho\left(\usurf\cdot\nabla\right)\left(\frac{1}{\varepsilon}\usurf\right) &= -\nabla\left(\varepsilon p\right) + \mu_e\nabla^2\usurf - \frac{\varepsilon\mu}{K}\usurf - \frac{\rho\varepsilon F_\varepsilon}{\sqrt{K}}\left|\usurf\right|\usurf
\end{aligned}
\end{equation}
and then is applied to 2D problems. Equation~\ref{eq:guo-incompressible} is declared in \cite{Guo2002} to derive from a theoretical analysis presented in \cite{Nithiarasu1997}. However, there are limits in this approach, and specifically: 
\begin{enumerate}
    \item[\emph{(i)}] the derivation is limited to 2D only;
    \item[\emph{(ii)}] the information pertaining to the compressible part of the target equation is not reported; 
    \item[\emph{(iii)}] the derivation is performed on a cubic REV and therefore lacks generality; 
    \item[\emph{(iv)}] the fields within the REV are implicitly considered to be approximable as a Taylor series truncated to the first order; and
    \item[\emph{(v)}] apart from the previous point, the assumptions and approximations being adopted during the derivation procedure are not clearly stated.
\end{enumerate} 

In \cite{Spaid1997,Krause2017,Trunk2018,Dapelo2019,Trunk2021,Marquardt2023}, no target equation in a parabolic form is reported. In \cite{Wang2015}, a target equation is derived extending the method of volume averaging proposed in \cite{whitakerMethodVolumeAveraging1999}, which represents the culmination of work carried out by Whitaker and collaborators for over thirty years  \citep{Whitaker1966,Howes1985,Whitaker1986,Quintard1994,Quintard1994a,Quintard1994b,Whitaker1996}. An equation is produced for the seepage velocity:
\begin{equation}\label{eq:guo-u}
\begin{aligned}
    \partial_t\rho + \nabla\cdot\left(\rho\useep\right) &= 0 , \\
    \partial_t\left(\rho\useep\right) + \nabla\cdot\left(\rho\useep\otimes\useep\right) &= -\nabla p + \mu_e\nabla^2\useep - \frac{\varepsilon\mu}{K}\useep - \frac{\rho\varepsilon^2 F_\varepsilon}{\sqrt{K}}\left|\useep\right|\useep
\end{aligned}
\end{equation}
and alternatively, for the surface velocity:
\begin{equation}\label{eq:guo-q}
\begin{aligned}
    \varepsilon\partial_t\rho + \nabla\cdot\left(\rho\usurf\right) &= 0 , \\
    \partial_t\left(\rho\usurf\right) + \nabla\cdot\left(\frac{\rho}{\varepsilon}\usurf\otimes\usurf\right) &= -\varepsilon\nabla p + \mu_e\nabla^2\usurf - \frac{\varepsilon\mu}{K}\usurf - \frac{\rho\varepsilon F_\varepsilon}{\sqrt{K}}\left|\usurf\right|\usurf .
\end{aligned}
\end{equation}
Contrary to the derivation of Equation~\ref{eq:guo-incompressible}, in Equations~\ref{eq:guo-u} and~\ref{eq:guo-q}: 
\begin{enumerate}
    \item[\emph{(i)}] the derivation is 3D; 
    \item[\emph{(ii)}] the equations account for (weak) compressibility;
    \item[\emph{(iii)}] a generic REV is considered; and
    \item[\emph{(iv)}] the fields are only assumed to be integrable, but no constraint about their Taylor series is considered. 
\end{enumerate}
Nevertheless, to the best of the authors' knowledge, the assumptions made when applying the volume-averaging procedure have not been reported in the literature. 
\subsection{Scope and structure of this work}
\label{s:scope}
The purpose of this work is to delineate the conditions under which the parabolic equations introduced in Section~\ref{s:parabolic} remain valid—and, by extension, to establish the domain of applicability for the corresponding Lattice-Boltzmann models.

In Section~\ref{s:nithiarasu}, we expand the procedure introduced in \cite{Nithiarasu1997} to the compressible 3D case, thereby showing that it is compatible with the work of \cite{Wang2015} and specifically, with Equation~\ref{eq:guo-q}.
Then, in Section~\ref{s:vol_avg}, we retrace in great detail the derivations of Equations~\ref{eq:guo-u} and~\ref{eq:guo-q} in order to explicitly list the assumptions and approximations underpinning the model, which were left implicit in \cite{Wang2015}. Such assumptions and approximations are then discussed  in Section~\ref{s:vol_avg}, and consolidated in Section~\ref{s:volAvg_assumption}.
Finally, in Section~\ref{s:multiphase} it is shown that the macroscopic description of porous media flow illustrated here (Equations~\ref{eq:guo-u} and~\ref{eq:guo-q}) can be interpreted as a particular case of multiphase flow whereby one phase is solid---in particular, a comparison is made with the volume-averaged Navier--Stokes equations for multiphase flow \citep{vanWachem2003}.
A discussion is presented in Section~\ref{s:discussion}. The conclusions follow in Section~\ref{s:conclusions}.

The main outcomes of this work are:
\begin{itemize}
    \item A list of assumptions underpinning Equations~\ref{eq:guo-u} and~\ref{eq:guo-q} (cf.\ Section~\ref{s:volAvg_assumption}):
    \begin {enumerate}
        \item Representative elementary volume;
        \item No large spatial deviations for all the fields\ldots
        \item \ldots which is reinforced to small deviations for the density field in order to account for quasi-incompressibility;
        \item Small Knudsen number;
        \item Porosity field non-zero at the REV scale and slowly varying in space and time at the macroscopic scale;
        \item Quasi-steadiness of the microscopic flow.
    \end{enumerate}
    \item An interpretation of the effective viscosity as the effect of non-negligible hydraulic dispersion, and its quantification in terms of a Smagorinsky-like viscosity correction (cf.\ Equation~\ref{eq:LES}):
\begin{equation}\label{eq:LES-anticipation}
    \mu_e\left(\left|\tens{S}\right|\right) = \mu + \varepsilon\left(\rho C_s\Delta\right)^2\left|\tens{S}\right|.
\end{equation}
    \item An interpretation of REV porous flow as a two-phase Euler--Euler two-component method, provided that the following assumptions are made (cf.\ Section~\ref{s:multiphase}:
    \begin{itemize}
        \item One phase is interpreted as the liquid phase, and the other one as the porous matrix (and not solved);
        \item Slowly-varying liquid phase fraction (\viz porosity);
        \item No interphase mass transfer;
        \item The interphase momentum transfer equates the Darcy and Forchheimer forces.
    \end{itemize}
\end{itemize}
\makeatletter
\g@addto@macro\input@path{{src/gener_nithiarasu/}}
\makeatother

\section{Generalization of the derivation in \cite{Nithiarasu1997}}
\label{s:nithiarasu}
\subsection{Representative elementary volume}
\label{s:Vrev}
The Representative Elementary Volume (REV) $\Vrev$ is defined as a volume element of linear scale $\lrev$, much larger than the linear length scale of the porous matrix elements $\lmat$, and much smaller than the size of the macroscopic system under analysis $\lsys$ \citep{Howes1985,whitakerMethodVolumeAveraging1999}:
\begin{equation}\label{eq:l_hierarchy}
\lmat \ll \lrev \ll \lsys . 
\end{equation}
In the simplified derivation of \cite{Nithiarasu1997}, $\Vrev$ is considered to be cubic. Here, for convenience, it is assumed to be centered at the spatial position $\vect{x}\equiv\left(x,\,y,\,x\right)$, \ie
\begin{equation}\label{eq:Vrev}
\Vrev = \left(x-\frac{\d x}{2},\,x+\frac{\d x}{2}\right)\times\left(y-\frac{\d y}{2},\,y+\frac{\d y}{2}\right)\times\left(z-\frac{\d z}{2},\,z+\frac{\d z}{2}\right) . 
\end{equation}

Let $\mathscr{A}$ be a unit of area of the order of magnitude $\mathscr{A} \sim \d x^2 \sim \d y^2 \sim \d z^2$. Let the average flow fraction be defined as
\begin{equation}\label{eq:a}
a := \frac{\text{average area available to the fluid flow within }\mathscr{A}}{\text{area of }\mathscr{A}} . 
\end{equation}
It is assumed that $a$ is constant or very slow-varying across the system's domain. Therefore, $a$ can be considered to be constant within $\Vrev$. It follows that, taking the $z$ direction as an example and without loss of generality:
\begin{equation}\label{eq:a_eps}
\varepsilon := \frac{\text{average volume occupied by the fluid within }\Vrev}{\text{volume of }\Vrev} = \frac{\dxy\int_{z-\frac{\dz}{2}}^{z+\frac{\dz}{2}}a\,\d\zeta}{\dxyz} = a
\end{equation}
because $a$ is an averaged quantity.
\subsection{Mass conservation}
\label{s:nithi_mass}
The conservation of the liquid phase mass is expressed as a balance between time variation of the liquid inside $\Vrev$ and the mass flux across the boundary $\partial\Vrev$. The mass of fluid inside $\Vrev$ can be approximated in terms of its density $\rho$ at $\vect{x}$ and on the time $t$, and the volume available to the liquid phase, $\varepsilon\,\dxyz$:
\begin{equation}\label{eq:nithi_M}
    M\left(t\right) = \left(\rho\varepsilon\right)\left(\vect{x},\,t\right)\,\dxyz . 
\end{equation}
The value of $M$ after a time $\dt$ is approximated as a Taylor series, truncated to the first order:
\begin{equation}\label{eq:nithi_M_tayor}
    M\left(t+\dt\right) \approx \left[\left(\rho\varepsilon\right)\left(\vect{x},\,t\right) + \frac{\partial\rho\varepsilon}{\partial t}\left(\vect{x},\,t\right)\dt\right]\,\dxyz . 
\end{equation}

The mass flux \eg across the positive $x$ direction is expressed in terms of the density at the centre of the face $yz$ of $\Vrev$ in the positive direction, the $x$ component of the seepage velocity, and the area available to flow $a\,\dyz\equiv\varepsilon\,\dyz$:
\begin{equation}\label{eq:nithi_PhiMx+}
    \Phi_x^+\left[M\left(t\right)\right] = \left(\rho u_x\varepsilon\right)\left(\vect{x}+\ivers\frac{\dx}{2},\,t\right)\,\dyz . 
\end{equation}
Approximating \(\Phi_x^+\) as a Taylor series, truncated to the first order yields
\begin{equation}\label{eq:nithi_PhiMx+taylor}
    \Phi_x^+\left[M\left(t\right)\right] \approx \left[\left(\rho u_x\varepsilon\right)\left(\vect{x},\,t\right) + \frac{\partial\rho u_x\varepsilon}{\partial x}\left(\vect{x},\,t\right)\frac{\dx}{2}\right]\,\dyz .
\end{equation}
The amount of fluid crossing the face $yz$ during the time $dt$ is approximated as $\Phi_x^+\left[M\right]\,\dt$.

Combining the foregoing, we obtain
\begin{equation}\label{eq:nithi_mass_u}
    \partial_t\left(\varepsilon\rho\right) + \nabla\cdot\left(\varepsilon\rho\useep\right)=0 ,
\end{equation}
or in terms of surface velocity:
\begin{equation}\label{eq:nithi_mass_q}
    \partial_t\left(\varepsilon\rho\right) + \nabla\cdot\left(\rho\usurf\right)=0 . 
\end{equation}
Equations~\ref{eq:nithi_mass_u} and~\ref{eq:nithi_mass_q} are compatible with the mass balances of Equations~\ref{eq:guo-u} and~\ref{eq:guo-q} respectively, provided that $\varepsilon$ is constant, or slowly varying, across the macroscopic size of the system.
\subsection{Momentum conservation}
\label{s:nithi_momentum}
The conservation of the liquid phase momentum is expressed as a balance between the time variation of the liquid momentum inside $\Vrev$, the momentum flux across the boundary $\partial\Vrev$, and the source/dissipation effects due to surface and body forces. In analogy to the case of the mass discussed in Section~\ref{s:nithi_mass}, the momentum \eg across the $y$-direction can be approximated in terms of its momentum density $\rho u_y$, \ie
\begin{equation}\label{eq:nithi_Py}
    P_y\left(t\right) = \left(\rho u_y\varepsilon\right)\left(\vect{x},\,t\right)\,\dxyz
\end{equation}
and its value after a time $\dt$ is again approximated as a Taylor series, truncated to the first order:
\begin{equation}\label{eq:nithi_Py_tayor}
    P_y\left(t+\dt\right) \approx \left[\left(\rho u_y\varepsilon\right)\left(\vect{x},\,t\right) + \frac{\partial\rho u_y\varepsilon}{\partial t}\left(\vect{x},\,t\right)\dt\right]\,\dxyz . 
\end{equation}

Again in analogy to the case of the mass discussed in Section~\ref{s:nithi_mass}, its flux \eg across the positive $x$-direction (through the face $yz$) is expressed as
\begin{equation}\label{eq:nithi_PhiPyx+}
    \Phi_x^+\left[P_y\left(t\right)\right] = \left(\rho u_y u_x\varepsilon\right)\left(\vect{x}+\ivers\frac{\dx}{2},\,t\right)\,\dyz 
\end{equation}
which is approximated as
\begin{equation}\label{eq:nithi_PhiPyx+taylor}
    \Phi_x^+\left[P_y\left(t\right)\right] \approx \left[\left(\rho u_y u_x\varepsilon\right)\left(\vect{x},\,t\right) + \frac{\partial\rho u_y u_x\varepsilon}{\partial x}\left(\vect{x},\,t\right)\frac{\dx}{2}\right]\,\dyz . 
\end{equation}
The amount of the momentum's $y$-component crossing the face $yz$ during the time $dt$ is approximated as $\Phi_x^+\left[P_y\right]\,\dt$.

The body force acting on $\Vrev$ can be approximated in terms of the body force density $\vect{b}$ at $\vect{x}$ and on the time $t$, and the volume available to the liquid phase, $\varepsilon\,\dxyz$. Concerning \eg the $x$-component, we have
\begin{equation}\label{eq:nithi_By}
    B_y\left(t\right) = \left(b_y\varepsilon\right)\left(\vect{x},\,t\right)\,\dxyz . 
\end{equation}

The surface force acting on $\partial\Vrev$ is separable into a transverse and a longitudinal contribution. The transverse contribution is related to the pressure. The contribution to the surface force coming from the pressure \eg on the positive $yz$-face is expressed in terms of the pressure at the center of the face $yz$ of $\Vrev$ in the positive direction, and the area available to flow $a\,\dyz\equiv\varepsilon\,\dyz$, \ie
\begin{equation}\label{eq:nithi_Pyz+}
    S_x^+\left(t\right) = \left(p\varepsilon\right)\left(\vect{x}+\ivers\frac{\dx}{2},\,t\right)\,\dyz \approx \left[\left(p\varepsilon\right)\left(\vect{x},\,t\right) + \frac{\partial p\varepsilon}{\partial x}\left(\vect{x},\,t\right)\frac{\dx}{2}\right]\,\dyz . 
\end{equation}
The transverse contribution is related to the shear stress. The contribution to the surface force coming from \eg the $y$-component of the shear stress acting on the positive $yz$-face $\tau_{xy}$ is expressed in terms of $\tau_{xy}$ evaluated at the centre of the positive face $yz$, and the area available to flow $a\,\dyz\equiv\varepsilon\,\dyz$. Thus
\begin{equation}\label{eq:nithi_Txy+}
    S_{xy}^+\left(t\right) = \left(\tau_{xy}\varepsilon\right)\left(\vect{x}+\ivers\frac{\dx}{2},\,t\right)\,\dyz \approx \left[\left(\tau_{xy}\varepsilon\right)\left(\vect{x},\,t\right) + \frac{\partial \tau_{xy}\varepsilon}{\partial x}\left(\vect{x},\,t\right)\frac{\dx}{2}\right]\,\dyz , 
\end{equation}
where $\tau_{xy}$ denotes the $x$-component of the shear stress along the $y$-direction and is approximated linearly as the difference between the values that $u_x$ takes at the extremes of $\Vrev$ along the $y$-direction, \ie
\begin{equation}\label{eq:nithi_shear}
    \tau_{xy}\left(\vect{x},\,t\right) \approx \mu\frac{u_x\left(\vect{x}+\jvers\frac{\dy}{2},\,t\right) - u_x\left(\vect{x}-\jvers\frac{\dy}{2},\,t\right)}{\dy} . 
\end{equation}

Combining the foregoing, we obtain
\begin{equation}\label{eq:nithi_momentum_u}
    \partial_t\left(\varepsilon\rho\useep\right) + \nabla\cdot\left(\varepsilon\rho\useep\otimes\useep\right) = -\nabla\left(\varepsilon p\right) + \mu\nabla^2\left(\varepsilon\useep\right) + \varepsilon\vect{b} , 
\end{equation}
or in terms of surface velocity:
\begin{equation}\label{eq:nithi_momentum_q}
    \partial_t\left(\rho\usurf\right) + \nabla\cdot\left(\frac{\rho}{\varepsilon}\usurf\otimes\usurf\right) = -\nabla\left(\varepsilon p\right) + \mu\nabla^2\usurf + \varepsilon\vect{b} . 
\end{equation}
Equations~\ref{eq:nithi_momentum_u} and~\ref{eq:nithi_momentum_q} are compatible with the momentum balances of Equations~\ref{eq:guo-u} and~\ref{eq:guo-q} respectively, provided that $\varepsilon$ is constant, or slowly varying, across the macroscopic size of the system, and that the body force is defined as
\begin{equation}\label{eq:nithi_b}
    \vect{b} = -\frac{\mu}{K}\useep - \frac{\rho\varepsilon F_\varepsilon}{\sqrt{K}}\left|\useep\right|\useep . 
\end{equation}
\makeatletter
\g@addto@macro\input@path{{src/vol_avg/}}
\makeatother

\section{Volume averaging procedure}
\label{s:vol_avg}
\subsection{Definitions and averaging theorems}
\label{s:volAvg_defs}
In the volume averaging procedure \citep{Whitaker1966,Howes1985,Whitaker1986,Quintard1994,Quintard1994a,Quintard1994b,Whitaker1996,whitakerMethodVolumeAveraging1999}, $\Vrev$ is  defined as in Section~\ref{s:Vrev}, and the hierarchy of length scales is the same as in Equation~\ref{eq:l_hierarchy}. However, the shape of $\Vrev$ is arbitrary---Equation~\ref{eq:Vrev} no longer holds.

Let $\Vrevf\subset\Vrev$ be the part of $\Vrev$ occupied by the fluid. Then, given an arbitrary field $\zeta$, the surface phase average of $\zeta$ \citep{Wang2015} is defined as
\begin{equation}\label{eq:avg}
    \avg{\zeta\left(\vect{x}\right)} = \frac{1}{\Vrev}\int\displaylimits_{\vect{x}+\vect{y}\in\Vrevf}\zeta\left(\vect{x}+\vect{y}\right)\,\d\vect{y}
\end{equation}
and the intrinsic phase average, as
\begin{equation}\label{eq:avgf}
    \avgf{\zeta\left(\vect{x}\right)} = \frac{1}{\Vrevf}\int\displaylimits_{\vect{x}+\vect{y}\in\Vrevf}\zeta\left(\vect{x}+\vect{y}\right)\,\d\vect{y} .
\end{equation}
Surface and intrinsic phase average are in fact convolutional products of a function $\zeta$ by indicator functions taking the values respectively of $1/\Vrev$ and $1/\Vrevf$ inside the domain of $\Vrevf$, and zero outside.
It can be immediately shown that
\begin{equation}\label{eq:epsilon_avg}
\frac{\avgf{\zeta}}{\avg{\zeta}} = \frac{\Vrevf}{\Vrev} \equiv \varepsilon \equiv \avg{1} .
\end{equation}
An averaged quantity is treated as a constant in successive averaging operations, \ie
\begin{equation}\label{eq:avg_of_avg}
\avg{\xi\avgf{\zeta}} = \avg{\xi}\avgf{\zeta}
\end{equation}
and in particular
\begin{equation}\label{eq:avg_of_avg1}
\avg{\avgf{\zeta}} = \avgf{\zeta}\avg{1} = \varepsilon\avgf{\zeta} .
\end{equation}
From the definitions of Equations~\ref{eq:avg} and~\ref{eq:avgf}, it follows that the seepage velocity can be interpreted as the intrinsic phase average of the microscopic velocity $\umat$, \ie
\begin{equation}\label{eq:def_u}
\useep \equiv \avgf{\umat}
\end{equation}
and that the surface velocity is identified as the surface phase average of the microscopic velocity, \ie
\begin{equation}\label{eq:def_q}
\usurf \equiv \avg{\umat} .
\end{equation}
Also, the liquid phase density $\rho$ can be interpreted as the intrinsic phase average of the microscopic liquid phase density $\rhomat$, \ie
\begin{equation}\label{eq:def_rho}
\rho \equiv \avgf{\rhomat}
\end{equation}
and because of Equation~\ref{eq:epsilon_avg}, we have
\begin{equation}\label{eq:def_rho_epsilon}
\avg{\rhomat} = \varepsilon\rho .
\end{equation}
This holds for pressure and body force as well, thus
\begin{align}
p \equiv \avgf{\pmat},\qquad\avg{\pmat} = \varepsilon p\label{eq:def_p},\\
\vect{b} \equiv \avgf{\bmat},\qquad\avg{\bmat} = \varepsilon\vect{b}\label{eq:def_b} .
\end{align}

Equation~\ref{eq:avgf} allows us to decompose the field $\zeta$ into its intrinsic phase average and its local spatial deviation $\dv{\zeta}$ as
\begin{equation}\label{eq:deviation}
    \zeta = \avgf{\zeta} + \dv{\zeta} .
\end{equation}
The definition of REV can be expressed mathematically via the requirement that the phase averages of the spatial deviation are zero:
\begin{equation}\label{eq:REV_in_maths}
    \avg{\dv{\zeta}} = \avgf{\dv{\zeta}} = 0 .
\end{equation}

The following averaging theorems apply:
\begin{equation}\label{eq:avg_dt}
\avg{\partial_t\zeta} = \partial_t\avg{\zeta} - \frac{1}{\Vrev}\intA{\zeta\useep_{\mathrm{surf}}\cdot\nvers} = \partial_t\avg{\zeta} ,
\end{equation}
where $\partial\Vrevf$ indicates the interfacial area between the fluid phase and the porous matrix within $\Vrev$. The integral in the equation above is zero because the velocity of the interface between the fluid phase and the solid matrix $\useep_{\mathrm{surf}}$ is zero as the porous matrix does not move.
Thus we obtain
\begin{equation}\label{eq:avg_grad}
\avg{\nabla\zeta} = \nabla\avg{\zeta} + \frac{1}{\Vrev}\intA{\zeta\nvers}
\end{equation}
and for a generic vector field:
\begin{equation}\label{eq:avg_div}
\avg{\nabla\cdot\vect{v}} = \nabla\cdot\avg{\vect{v}} + \frac{1}{\Vrev}\intA{\vect{v}\cdot\nvers} .
\end{equation}

\makeatletter
\g@addto@macro\input@path{{src/vol_avg/equations/}}
\makeatother

\subsection{Detailed derivation of the field equations}
\label{s:volAvg_equations}
A detailed derivation of the Equations~\ref{eq:guo-u} and~\ref{eq:guo-q} is given below, following \cite{Wang2015}, but making explicit mention of the underlying assumption of every step. A summary of the assumptions is given in Section~\ref{s:volAvg_assumption}.

In addition to the mathematical definition of REV presented in Equation~\ref{eq:REV_in_maths}, it is further assumed that the spatial deviation of a generic field $\zeta$ is no larger, in order of magnitude, than its intrinsic phase average, \ie
\begin{equation}\label{eq:order_field}
    \left|\dv{\zeta}\right| \approx \order{\avgf{\zeta}} .
\end{equation}
It follows that
\begin{equation}\label{eq:order_field_time_derivatives}
    \left|\partial_t\dv{\zeta}\right| \approx \order{\frac{\avgf{\zeta}}{\tmat}}
\end{equation}
and
\begin{equation}\label{eq:order_field_spatial_derivatives}
    \left|\nabla\dv{\zeta}\right| \approx \order{\frac{\avgf{\zeta}}{\lmat}}\quad\text{and}\quad\left|\nabla\avgf{\zeta}\right| \approx \order{\frac{\avgf{\zeta}}{\lsys}} .
\end{equation}
Because of the hierarchy of scales of Equation~\ref{eq:l_hierarchy}, it follows that
\begin{equation}\label{eq:order_field_spatial_derivatives_compare}
    \left|\nabla\dv{\zeta}\right| \gg \left|\nabla\avgf{\zeta}\right| .
\end{equation}

\subsubsection{Mass balance}
\label{s:volAvg_mass}
The surface volume average of the microscopic mass equation reads
\begin{equation}\label{eq:mass_micro}
    \avg{\partial_t\rhomat} + \avg{\nabla\cdot\left(\rhomat\umat\right)} = 0 .
\end{equation}

Because of the averaging theorem of Equation~\ref{eq:avg_dt} and Equations~\ref{eq:def_rho} and~\ref{eq:def_rho_epsilon}, we have
\begin{equation}\label{eq:mass_dt}
    \avg{\partial_t\rhomat} = \partial_t\avg{\rhomat} = \partial_t\left(\varepsilon\rho\right) .
\end{equation}

No-slip boundary conditions are assumed, \ie
\begin{equation}\label{eq:noSlip}
    \umat = 0\;\text{ at }\partial\Vrevf\qquad\text{or}\qquad\useep = -\dv{\umat}\;\text{ at }\partial\Vrevf .
\end{equation}
This is the case if the Knudsen number, defined as the ratio between molecular mean free path and $\lmat$, is small:
\begin{equation}\label{eq:Kn}
    \Kn \ll 1
\end{equation}
Therefore, from the averaging theorem of Equation~\ref{eq:avg_div}, we obtain
\begin{equation}\label{eq:mass_div1}
\avg{\nabla\cdot\left(\rhomat\umat\right)} = \nabla\cdot\avg{\rhomat\umat} + \frac{1}{\Vrev}\intA{\rhomat\umat\cdot\nvers} = \nabla\cdot\avg{\rhomat\umat} .
\end{equation}
Furthermore, the terms in $\avg{\rhomat\umat}$ can be decomposed into intrinsic phase average and deviation, \ie
\begin{equation}\label{eq:mass_div2}
\begin{aligned}
\avg{\rhomat\umat} &= \avg{\left(\avgf{\rhomat}+\dv{\rhomat}\right)\left(\avgf{\umat}+\dv{\umat}\right)} \simeq \avg{\avgf{\rhomat}\left(\avgf{\umat}+\dv{\umat}\right)} \\
&=\avg{\avgf{\rhomat}\avgf{\umat}} + \avg{\avgf{\rhomat}\dv{\umat}} . 
\end{aligned}
\end{equation}
The terms in $\dv{\rhomat}$ can be ignored if quasi-incompressibility is assumed, which means, in this case, a more stringent version of Equation~\ref{eq:order_field} for the density:
\begin{equation}\label{eq:mass_noPressureWaves}
    \dv{\rhomat} \ll \rho .
\end{equation}
Furthermore, from Equation~\ref{eq:avg_of_avg1}, we have
\begin{equation}\label{eq:mass_div3}
\avg{\avgf{\rhomat}\avgf{\umat}} = \avgf{\rhomat}\avgf{\umat}\avg{1} = \varepsilon\avgf{\rhomat}\avgf{\umat} \equiv \varepsilon\rho\useep
\end{equation}
and similarly
\begin{equation}\label{eq:mass_div4}
    \avg{\avgf{\rhomat}\dv{\umat}} = \avg{\avgf{\rhomat}}\dv{\umat} = 0
\end{equation}
because of Equation~\ref{eq:REV_in_maths}.

Combining the foregoing, we obtain
\begin{equation}\label{eq:mass_final}
    \partial_t\left(\varepsilon\rho\right) + \nabla\cdot\left(\varepsilon\rho\useep\right) = 0 .
\end{equation}
The equation above reduces to the mass balances proposed by \cite{Wang2015} (Equations~\ref{eq:guo-u} and~\ref{eq:guo-q}, first parts) if the porosity field is assumed to be slowly varying in space and time:
\begin{equation}\label{eq:mass_assump_epsilon}
    \rho\partial_t\varepsilon \ll \varepsilon\partial_t\rho , \qquad \rho\useep\cdot\nabla\varepsilon \ll \varepsilon\nabla\cdot\left(\rho\useep\right) .
\end{equation}
\subsubsection{Momentum balance}
\label{s:volAvg_momentum}
The surface volume average of the microscopic momentum equation reads
\begin{equation}\label{eq:momentum_micro}
    \avg{\partial_t\left(\rhomat\umat\right)} + \avg{\nabla\cdot\left(\rhomat\umat\otimes\umat\right)} = -\avg{\nabla\pmat} + \mu\avg{\nabla^2\umat} + \avg{\bmat} . 
\end{equation}

We immediately have $\avg{\bmat} = \varepsilon\vect{b}$ from Equation~\ref{eq:def_b}. We have $\avg{\partial_t\left(\rhomat\umat\right)} = \partial_t\avg{\rhomat\umat}$ from the averaging theorem of Equation~\ref{eq:avg_dt}. Then, decomposing into intrinsic phase average and deviation, and following the same steps of Equations~\ref{eq:mass_div2}, \ref{eq:mass_div3} and~\ref{eq:mass_div4}, we obtain
\begin{equation}\label{eq:momentum_dt}
    \partial_t\avg{\rhomat\umat} = \partial_t\left(\varepsilon\rho\useep\right) + \partial_t\avg{\dv{\rhomat}\dv{\umat}} \simeq \varepsilon\partial_t\left(\rho\useep\right) + \partial_t\avg{\dv{\rhomat}\dv{\umat}} . 
\end{equation}
If a slowly-varying porosity is assumed, the equation above transforms to 
\begin{equation}\label{eq:momentum_assump_epsilon_dt}
    \rho\useep\partial_t\varepsilon \ll \varepsilon\partial_t\left(\rho\useep\right) = \varepsilon\rho\partial_t\useep + \varepsilon\useep\partial_t\rho , 
\end{equation}
which is automatically satisfied if the first of Equation~\ref{eq:mass_assump_epsilon} holds.

In analogy to Equation~\ref{eq:mass_div1}, because of the averaging theorem of Equation~\ref{eq:avg_div} and the non-slip boundary conditions, we have $\avg{\nabla\cdot\left(\rhomat\umat\otimes\umat\right)} = \nabla\cdot\avg{\rhomat\umat\otimes\umat}$. Then, following once more the same steps of Equations~\ref{eq:mass_div2}, \ref{eq:mass_div3} and~\ref{eq:mass_div4}, we have
\begin{equation}\label{eq:momentum_div1}
\begin{aligned}
\nabla\cdot\avg{\rhomat\umat\otimes\umat} &\simeq \nabla\cdot\avg{\avgf{\rhomat}\left(\avgf{\umat}+\dv{\umat}\right)\otimes\left(\avgf{\umat}+\dv{\umat}\right)}\\
&= \nabla\cdot\left(\varepsilon\rho\useep\otimes\useep\right) + \nabla\cdot\left(\rho\avg{\dv{\umat}\otimes\dv{\umat}}\right) .
\end{aligned}
\end{equation}
The assumption of a slowly-varying porosity field means the following condition in addition to the ones expressed in Equations~\ref{eq:mass_assump_epsilon} and~\ref{eq:momentum_assump_epsilon_dt}:
\begin{equation}\label{eq:momentum_assump_epsilon_div}
    \left(\nabla\varepsilon\right)\cdot\left(\rho\useep\otimes\useep\right) \ll \varepsilon\nabla\cdot\left(\rho\useep\otimes\useep\right) , 
\end{equation}
which is automatically satisfied if the second of Equation~\ref{eq:mass_assump_epsilon} holds.
This leads to
\begin{equation}\label{eq:momentum_div2}
\nabla\cdot\avg{\rhomat\umat\otimes\umat} = \varepsilon\nabla\cdot\left(\rho\useep\otimes\useep\right) + \nabla\cdot\left(\rho\avg{\dv{\umat}\otimes\dv{\umat}}\right) .
\end{equation}

Applying the averaging theorem of Equation~\ref{eq:avg_grad} to the pressure term, we have
\begin{equation}\label{eq:momentum_p1}
    \avg{\nabla\pmat} = \nabla\avg{\pmat} + \frac{1}{\Vrev}\intA{\pmat\nvers} .
\end{equation}
Then, separating intrinsic phase average and spatial deviation gives
\begin{equation}\label{eq:momentum_p2}
\begin{aligned}
    \avg{\nabla\pmat} &= \nabla\avg{\avgf{\pmat}+\dv{\pmat}} + \frac{1}{\Vrev}\intA{\left(\avgf{\pmat}+\dv{\pmat}\right)\nvers}\\
    &= \nabla\left(\varepsilon p\right) + \nabla 0 + \frac{p}{\Vrev}\intA{\nvers} + \frac{1}{\Vrev}\intA{\dv{\pmat}\nvers} .
\end{aligned}
\end{equation}
A result from the distribution theory is \citep{Quintard1994a}
\begin{equation}\label{eq:nabla_epsilon}
    \frac{1}{\Vrev}\intA{\nvers} = -\nabla\varepsilon
\end{equation}
and therefore
\begin{equation}\label{eq:momentum_p3}
    \avg{\nabla\pmat} = \varepsilon\nabla p + \frac{1}{\Vrev}\intA{\dv{\pmat}\nvers} .
\end{equation}

Applying the averaging theorem of Equation~\ref{eq:avg_div} to the Laplacian term $\avg{\nabla^2\umat}\equiv\avg{\nabla\cdot\left(\nabla\otimes\umat\right)}$ yields
\begin{equation}\label{eq:momentum_dissip1}
    \avg{\nabla^2\umat} = \nabla\cdot\avg{\nabla\otimes\umat} + \frac{1}{\Vrev}\intA{\left(\nabla\otimes\umat\right)\cdot\nvers} .
\end{equation}
Separating intrinsic phase average and spatial deviation inside the area integral and applying Equation~\ref{eq:nabla_epsilon}, we have that
\begin{equation}\label{eq:momentum_dissip2}
    \avg{\nabla^2\umat} = \nabla\cdot\avg{\nabla\otimes\umat} - \left(\nabla\varepsilon\right)\cdot\left(\nabla\otimes\useep\right) + \frac{1}{\Vrev}\intA{\left(\nabla\otimes\dv{\umat}\right)\cdot\nvers} .
\end{equation}
Applying the averaging theorem of Equation~\ref{eq:avg_grad} to the first term on the right-hand side yields
\begin{equation}\label{eq:momentum_dissip3}
    \avg{\nabla\otimes\umat} = \nabla\otimes\avg{\umat} + \frac{1}{\Vrev}\intA{\umat\otimes\nvers} = \nabla\otimes\avg{\umat}
\end{equation}
because of the no-slip boundary conditions (Equation~\ref{eq:noSlip}. Therefore
\begin{equation}\label{eq:momentum_dissip4}
    \avg{\nabla^2\umat} = \nabla^2\usurf - \left(\nabla\varepsilon\right)\cdot\left(\nabla\otimes\useep\right) + \frac{1}{\Vrev}\intA{\left(\nabla\otimes\dv{\umat}\right)\cdot\nvers} .
\end{equation}
In terms of the seepage velocity only, we have
\begin{equation}\label{eq:momentum_dissip5}
    \avg{\nabla^2\umat} = \varepsilon\nabla^2\useep + \useep\nabla^2\varepsilon - \left(\nabla\varepsilon\right)\cdot\left(\nabla\otimes\useep\right) + \frac{1}{\Vrev}\intA{\left(\nabla\otimes\dv{\umat}\right)\cdot\nvers} .
\end{equation}
The assumption of a slowly-varying porosity field means the following condition in addition to the ones expressed in Equations~\ref{eq:mass_assump_epsilon}, \ref{eq:momentum_assump_epsilon_dt} and~\ref{eq:momentum_assump_epsilon_div}:
\begin{equation}\label{eq:momentum_assump_epsilon_Laplacian}
    \left(\nabla\varepsilon\right)\cdot\left(\nabla\otimes\useep\right) \ll \varepsilon\nabla^2\useep , \qquad \useep\nabla^2\varepsilon \ll \varepsilon\nabla^2\useep .
\end{equation}

Assuming that the porosity is non-zero at the REV scale everywhere within the system domain and dividing by $\varepsilon$ allows us to obtain a filtered equation for the momentum balance at the REV scale, whereby ``filter'' means that the information about the microscopic features of the flow is filtered out by closures \citep{Whitaker1996}, \ie
\begin{equation}\label{eq:momentum_before_closures}
\begin{aligned}
    \partial_t\left(\rho\useep\right) + \nabla\cdot\left(\rho\useep\otimes\useep\right) + \filterV =
    -\nabla p + \mu\nabla^2\useep + \frac{\mu}{\varepsilon}\useep\nabla^2\varepsilon - \frac{\mu}{\varepsilon}\left(\nabla\varepsilon\right)\cdot\left(\nabla\otimes\useep\right) + \filterA , 
\end{aligned}
\end{equation}
where $\filterV$ and $\filterA$ are respectively a volume and a surface (convolutional) filter \citep{Whitaker1996}, and are defined as
\begin{align}
    \filterV &= \frac{1}{\varepsilon}\partial_t\avg{\dv{\rhomat}\dv{\umat}} + \frac{1}{\varepsilon}\nabla\cdot\left(\rho\avg{\dv{\umat}\otimes\dv{\umat}}\right)\label{eq:momentum_filterV} , \\
    \filterA &= \frac{1}{\Vrevf}\intA{\left(-\tens{I}\dv{\pmat} + \mu\nabla\otimes\dv{\umat}\right)\cdot\nvers}\label{eq:momentum_filterA} .
\end{align}
\subsubsection{Closures for the momentum balance}
\label{s:closures}
Concerning the area filter $\filterA$ (Equation~\ref{eq:momentum_filterA}), it is shown in \cite{Whitaker1996} that, upon assuming that the microscopic flow is quasi-steady,
\begin{equation}\label{eq:area_quasi_steady}
    \partial_t\left(\rho\dv{\umat}\right) \ll \mu\nabla^2\dv{\umat}
\end{equation}
can be reduced to an expression of the permeability tensor $\tens{K}$ and an inertial term $\tens{F}$ proportional to the microscopic Reynolds number $\Remat$:
\begin{equation}\label{eq:filterA_closed}
    \filterA = -\mu\tens{K}^{-1}\cdot\left[\tens{I} + \tens{F}\left(\Remat\right)\right]\cdot\useep .
\end{equation}
Under isotropic conditions, $\tens{K}$ and $\tens{F}$ become scalars and Equation~\ref{eq:filterA_closed} reduces to the forms proposed by \cite{Wang2015} (Equations~\ref{eq:guo-u} and~\ref{eq:guo-q}, second lines):
\begin{equation}\label{eq:e:filterA_closed_simp}
    \filterA = -\frac{\mu}{K}\useep - \frac{\rho\varepsilon F_\varepsilon}{\sqrt{K}}\left|\useep\right|\useep .
\end{equation}

Concerning the volume filter $\filterV$ (Equation~\ref{eq:momentum_filterV}), the term $\nabla\cdot\left(\rho\avg{\dv{\umat}\otimes\dv{\umat}}\right)$ or ``hydraulic dispersion'' is investigated in both \cite{Wang2015} and \cite{Whitaker1996}, whereas it is omitted altogether in \cite{Wood2013,Lasseux2019} due to the linearity of their models. \cite{Wang2015} neglect the hydraulic dispersion confirming that this follows from the assumption of $\dv{\umat}\ll\useep$, citing \cite{Hsu1990}. However, in \cite{Hsu1990}, it is assumed that only the hydraulic dispersion, not the spatial deviation of the velocity, is negligible. Indeed, whilst a negligible spatial deviation of the velocity implies a negligible hydraulic dispersion, \viz $\dv{\useep}\ll\useep \Rightarrow \nabla\cdot\left(\rho\avg{\dv{\umat}\otimes\dv{\umat}}\right)\ll\nabla\cdot\left(\rho\avg{\useep\otimes\useep}\right)$, the converse is not true. In fact, the assumption $\dv{\umat}\ll\useep$ cannot be correct as we have $\dv{\umat}=-\useep$ at $\partial\Vrevf$ because of the no-slip boundary conditions (Equation~\ref{eq:noSlip}). In \cite{whitakerMethodVolumeAveraging1999} and preceding papers, no explicit assumption is made on the hydraulic dispersion: it is simply neglected upon the observation that the Laplacian part of Equation~\ref{eq:momentum_before_closures}, or Brinkman correction, is much smaller than $\filterA$---and therefore, the hydraulic dispersion as well because of the condition of Equation~\ref{eq:order_field}. Considering that, in this work, the Brinkman correction is not neglected, the following reformulation of the assumption made in \cite{Hsu1990} would appear to be the most appropriate:
\begin{equation}\label{eq:hydraulic_dispersion}
    \nabla\cdot\left(\rho\avg{\dv{\umat}\otimes\dv{\umat}}\right)\ll\nabla\cdot\left(\rho\useep\otimes\useep\right) .
\end{equation}
However, no justification for this assumption is provided in \cite{Hsu1990}. As such, to our understanding, a deviation from this assumption remains an open possibility. Such deviation may be modelled mathematically by following the steps of the RANS theory and introducing a Reynolds-stress-like tensor
\begin{equation}\label{eq:RANS}
    \tens{R} \equiv \varepsilon\rho\avg{\dv{\umat}\otimes\dv{\umat}} .
\end{equation}
It should be noted that the assumption of equation~\ref{eq:momentum_assump_epsilon_Laplacian} prevents the emergence of extra terms in $\varepsilon$ from taking the divergence of the term above. Alternatively, the homogeneous LES theory can be followed, introducing a Smagorinsky-like constant and substituting it in the Brinkman correction, \ie
\begin{equation}\label{eq:LES}
    \mu \longrightarrow \mu_e\left(\left|\tens{S}\right|\right) = \mu + \varepsilon\left(\rho C_s\Delta\right)^2\left|\tens{S}\right|,
\end{equation}
where $\tens{S}=\left[\left(\nabla\otimes\useep\right)+\left(\nabla\otimes\useep\right)^{\mathrm{T}}\right]/2$ is the shear rate. This also offers the possibility to explain the discrepancy between viscosity and Brinkman's effective viscosity.

The term $\partial_t\avg{\dv{\rhomat}\dv{\umat}}/\varepsilon$ in $\filterV$ was not investigated in \cite{whitakerMethodVolumeAveraging1999} and preceding papers nor in \cite{Hsu1990}, whereas it is shown in \cite{Wood2013,Lasseux2019} to induce a memory effect. However, it can be omitted under the assumption of quasi-compressibility (Equation~\ref{eq:mass_noPressureWaves}). In this context, this hypothesis implies that
\begin{equation}\label{eq:volume_quasi_steady}
    \partial_t\avg{\dv{\rhomat}\dv{\umat}} \ll \nabla\cdot\left(\rho\avg{\dv{\umat}\otimes\dv{\umat}}\right) .
\end{equation}
To prove the equation above, we observe that, due to Equation~\ref{eq:order_field_spatial_derivatives},
\begin{equation}\label{eq:volume_quasi_steady_dim1}
    \nabla\cdot\left(\rho\avg{\dv{\umat}\otimes\dv{\umat}}\right) \approx \rho\frac{\dv{\mat{u}}^2}{\lmat}
\end{equation}
and because
\begin{equation}\label{eq:dim_u}
\dv{\mat{u}}\approx\frac{\lmat}{\tmat} , 
\end{equation}
we obtain that
\begin{equation}\label{eq:volume_quasi_steady_dim2}
    \nabla\cdot\left(\rho\avg{\dv{\umat}\otimes\dv{\umat}}\right) \approx \rho\frac{\lmat}{\tmat^2} .
\end{equation}
Similarly, due to Equation~\ref{eq:order_field_time_derivatives}, we have
\begin{equation}\label{eq:volume_quasi_steady_dim3}
    \partial_t\avg{\dv{\rhomat}\dv{\umat}} \approx \frac{\dv{\rhomat}\dv{\mat{u}}}{\tmat} \ll\frac{\rho\dv{\mat{u}}}{\tmat} ,
\end{equation}
where the last inequality is due to Equation~\ref{eq:mass_noPressureWaves}. Thus
\begin{equation}\label{eq:volume_quasi_steady_dim4}
    \partial_t\avg{\dv{\rhomat}\dv{\umat}} \ll\rho\frac{\lmat}{\tmat^2} \approx \nabla\cdot\left(\rho\avg{\dv{\umat}\otimes\dv{\umat}}\right) .
\end{equation}
These observations align with the discussion in \cite{Lasseux2019}, which notes that the memory term becomes negligible under steady (or quasi-steady) forcing and in the creeping-flow regime. In heuristic terms, memory effects vanish once the microscopic relaxation time is much shorter than the characteristic timescale of macroscopic flow variations.
\subsection{Summary of the assumptions}
\label{s:volAvg_assumption}
The assumptions being formulated across Section~\ref{s:vol_avg} are provided below.
\begin{enumerate}[leftmargin=*, label=\textbf{Assumption \arabic*:}, ref=Assumption~\arabic*]
\item Representative Elementary Volume (Equations~\ref{eq:l_hierarchy} and~\ref{eq:REV_in_maths}):
\begin{align}
& \lmat \ll \lrev \ll \lsys\label{eq:assumpt_l_hierarchy}, \\
& \avg{\dv{\zeta}} = \avgf{\dv{\zeta}} = 0\label{eq:assumpt_REV_in_maths} .\\
\end{align}
\item No large spatial deviations (Equations~\ref{eq:order_field}):
\begin{equation}\label{eq:assumpt_order_field}
    \left|\dv{\zeta}\right| \approx \order{\avgf{\zeta}} .
\end{equation}
\item The assumption above is modified in order to impose quasi-incompressibility (Equation~\ref{eq:mass_noPressureWaves}):
\begin{equation}\label{eq:assumpt_mass_noPressureWaves}
    \dv{\rhomat} \ll \rho .
\end{equation}
\item Small Knudsen number (Equation~\ref{eq:Kn}):
\begin{equation}\label{eq:assumpt_Kn}
    \Kn \ll 1 .
\end{equation}
\item Porosity field non-zero at the REV scale and slowly-varying on space and time (Equations~\ref{eq:mass_assump_epsilon}) and~\ref{eq:momentum_assump_epsilon_Laplacian}:
\begin{equation}\label{eq:assump_epsilon}
\begin{aligned}
    \varepsilon &> 0 , \\
    \rho\partial_t\varepsilon &\ll \varepsilon\partial_t\rho , \\
    \rho\useep\cdot\nabla\varepsilon &\ll \varepsilon\nabla\cdot\left(\rho\useep\right) , \\
    \left(\nabla\varepsilon\right)\cdot\left(\nabla\otimes\useep\right) &\ll \varepsilon\nabla^2\useep , \\
    \useep\nabla^2\varepsilon &\ll \varepsilon\nabla^2\useep .
\end{aligned}
\end{equation}
\item Quasi-steadiness of the microscopic flow (Equation~\ref{eq:area_quasi_steady}):
\begin{equation}\label{eq:assumpt_momentum_filterA}
    \partial_t\left(\rho\dv{\umat}\right) \ll \mu\nabla^2\dv{\umat} .
\end{equation}
Following Equations~\ref{eq:order_field_time_derivatives} and~\ref{eq:order_field_spatial_derivatives}, this is equivalent to
\begin{equation}\label{eq:assumpt_momentum_filterA_dim}
    \frac{\tmat}{\tsys} \gg \left(\frac{\lmat}{\lsys}\right)^2 \approx \frac{\lmat}{\lsys} .
\end{equation}
\item \textbf{(Optional)} Negligible hydraulic dispersion (Equation~\ref{eq:hydraulic_dispersion}):
\begin{equation}\label{eq:assumpt_hydraulic_dispersion}
    \nabla\cdot\left(\rho\avg{\dv{\umat}\otimes\dv{\umat}}\right)\ll\nabla\cdot\left(\rho\useep\otimes\useep\right) .
\end{equation}
This is equivalent to assume small $\Remat$. If this assumption is not met, then a Reynolds-stress-like tensor (Equation~\ref{eq:RANS}) needs to be modelled---or alternatively, a shear-stress-dependent effective viscosity must be introduced as in Equation~\ref{eq:LES}, \ie
\begin{equation}\label{eq:assumpt_LES}
    \mu \longrightarrow \mu_e\left(\left|\tens{S}\right|\right) = \mu + \varepsilon\left(\rho C_s\Delta\right)^2\left|\tens{S}\right| .
\end{equation}
\end{enumerate}
\section{Comparision with the multiphase volume-averaging}
\label{s:multiphase}
It has been pointed out in \citep{Dapelo2025} that the REV model defined by Equations~\ref{eq:guo-u} or~\ref{eq:guo-q} can be interpreted as an Euler--Euler multiphase (or more appropriately, multicomponent) model wherein one phase (component) represents the rigid porous matrix, and the other one, the liquid saturating it. As such, only the liquid component of the multiphase model is solved. In this section, we evidence the substantial identity between the volume-averaging approaches adopted in the REV (specifically, \cite{whitakerMethodVolumeAveraging1999} and preceding papers), and in multiphase modelling; and we state the conditions under which the two modelling approaches are identical.

The Euler--Euler framework in Computational Fluid Dynamics (EE CFD) models each phase of a multiphase flow as an interpenetrating continuum, enabling the simulation of complex flows without the need to track individual interfaces. Closure relations specific to the particular model being considered are provided to model interphase momentum and mass (in case \eg of reacting species) transfer. The macroscopic equations of motion are obtained by applying the volume-averaging procedure described in \cite{vanWachem2003, Yeoh2019}, to the microscopic Navier--Stokes equations.
 Given a control volume $V$ much smaller than the scale at which the macroscopic details are resolved and with centroid $\vect{x}_V$, the volume $V^p$ occupied by the phase $p$ is equal to the integral of a phase indicator function, defined as equal to $1$ if a point is occupied by the phase $p$, and 0 otherwise, \ie
\begin{equation}\label{eq:Vp}
  V^p = \int\limits_V\chi^p\left(\vect{x}\right)\,\d\vect{x} .
\end{equation}
The volume fraction is defined as
\begin{equation}\label{eq:alpha_p}
  \alpha^p = \frac{V^p}{V} .
\end{equation}
The phase-averaging of a generic quantity $\zeta$ is defined as
\begin{equation}\label{eq:phase_avg}
  \zeta^p\left(\vect{x}_V\right) = \frac{1}{V^p}\int\limits_V \zeta\left(\vect{x}\right)\,\chi^p\left(\vect{x}\right)\,\d\vect{x} .
\end{equation}
The identity between the volume-averaging procedure for EE described above, and the one for REV porous flow reported in Section~\ref{s:vol_avg}, is noted by restricting $p$ to $p\in\left\{L,\,S\right\}$ where $L$ represents liquid and $S$ represents solid, and noting the following:
\begin{itemize}
    \item $V$ in EE corresponds to $\Vrev$ in REV porous flow;
    \item $V^L$ in EE corresponds to $\Vrevf$ in REV porous flow;
    \item Equation~\ref{eq:phase_avg} in EE corresponds to the intrinsic phase average (Equation~\ref{eq:avgf}) in REV porous flow;
    \item $\alpha~L$ in EE corresponds to $\varepsilon$ in REV porous flow.
\end{itemize}

The macroscopic mass balance for the EE is
\begin{equation}\label{eq:macro_mass}
  \partial_t\left(\alpha^p\rho^p\right) + \nabla\cdot\left(\alpha^p\rho^p\vect{u}^p\right) = \sum_q\left(\dot{m}^{qp} - \dot{m}^{pq}\right) . 
\end{equation}
The momentum balance reads \citep{Yeoh2018}
\begin{equation}\label{eq:macro_momentum_multipressure}
\begin{aligned}
  \partial_t\left(\alpha^p\rho^p\vect{u}^p\right) + \nabla\cdot\left(\alpha^p\rho^p\vect{u}^p\otimes\vect{u}^p\right) = 
  -\nabla\left(\alpha^p p^p\right) + \nabla\cdot\left(\alpha^p\mu\tens{S}^p\right)  + \alpha^p\rho^p\vect{b}^p + \sum_q\left(\vect{M}^{qp} - \vect{M}^{pq}\right) .
\end{aligned}
\end{equation}
In Finite-Volume algorithms, the equation above is altered by using the same pressure $p^p\equiv p$ for all the phases, in order to provide much simpler numerical implementations \citep{Dapelo2025}. This is equivalent to neglecting the fluctuations of the dispersed phase into the pressure balance of the continuous one in the case of dispersed flows \citep{Zhang2007}.
The correspondence between the EE mass and momentum balance equations described above, and the intrinsic-phase-averaged equations for REV porous flow (Equation~\ref{eq:guo-u}) is recovered provided that;
\begin{itemize}
    \item The solution procedure in Equations~\ref{eq:macro_mass} and~\ref{eq:macro_momentum_multipressure} is restricted to the liquid phase $p\equiv L$ and omitted for the solid matrix $p\equiv S$;
    \item Slowly-varying liquid phase fraction (\viz porosity) is assumed;
    \item No interphase mass transfer occurs;
    \item The interphase momentum transfer equates the Darcy and Forchheimer forces (reported \eg in Equation~\ref{eq:nithi_b}).
\end{itemize}
\section{Discussion}
\label{s:discussion}
The following limitations were found in \cite{Nithiarasu1997}'s procedure, underpinning Equation~\ref{eq:guo-incompressible}:
\begin{enumerate}
    \item\label{i:2d} The derivation is limited to 2D only;
    \item\label{i:compr} The information pertaining the compressible part of the target equation is not reported; 
    \item\label{i:cubicREV} The derivation is performed on a cubic REV and therefore lacks of generality;
    \item\label{i:taylor} The fields within the REV are implicitly considered to be approximable as a Taylor series truncated to the first order;
    \item\label{i:assumptions} Notwithstanding the previous point, the assumptions and approximations being adopted during the derivation procedure are not clearly stated.
\end{enumerate}
The procedure is reviewed and expanded in Section~\ref{s:nithiarasu}. The expanded procedure overcomes limits~\ref{i:2d} and~\ref{i:compr}, whereas the other three persist. The resulting balance equations are shown to be compatible with Equations~\ref{eq:guo-u} and~\ref{eq:guo-q}, which are introduced in \cite{Wang2015}. Therefore, the expanded procedure of Section~\ref{s:nithiarasu} is proven to be compatible with, but not equivalent to, the one of \cite{Wang2015}.

The procedure proposed by \cite{Wang2015}, which expands the work presented in \cite{whitakerMethodVolumeAveraging1999} and preceding papers and underpins Equations~\ref{eq:guo-u} and~\ref{eq:guo-q}, is free from the limits listed above, bar limit~\ref{i:assumptions}. The procedure is reviewed in Section~\ref{s:vol_avg}, and limit~\ref{i:assumptions} is overcome---specifically, the underlying assumptions and approximations of the model are listed in Section~\ref{s:volAvg_assumption}.

During the analysis of the limits and limitations, it is noted in Section~\ref{s:closures} that the assumption of negligible hydraulic dispersion commonly assumed in the literature is not justifiable by clear arguments and therefore may be inappropriate for some problems. In fact, not making this assumption can offer a quantitative and qualitative description of the discrepancy between viscosity and effective viscosity in the Brinkman correction through Equation~\ref{eq:assumpt_LES}.
\section{Conclusions}
\label{s:conclusions}
Parabolic models for fluid flow in REV porous media are of interest because, among other things, they offer viable macroscopic target equations for Lattice-Boltzmann models.
Here, it is shown that a previous result \citep{Nithiarasu1997} is compatible with a more recent model based on the theory of volume averaging \citep{Wang2015}.

The limits and assumptions of parabolic equations---and therefore, of Lattice-Boltzmann models---for flow in REV porous media are investigated and listed: (1) clear separation of scales between microscopic, REV nad macroscopic length; (2) no large spatial deviations for the macroscopic fields, and in particular (3) small spatial deviations for the density; (4) small Knudsen number; (5) porosity field non-zero at the REV scale and slowly-varying on space and time; and (6) quasi-steadiness of the microscopic flow.

It is shown that the commonly adopted assumption of negligible hydraulic dispersion is not justifiable by a clear argument. Conversely, assuming non-negligible hydraulic dispersion can explain the discrepancy between viscosity and effective viscosity in the Brinkman correction--for instance, by modelling the correction as a Smagorinsky-like term proportional to the shear rate.

Finally, conditions are listed under which the volume-averaged model for flow in REV porous media is equivalent to an EE two-component model: (a) identification of one phase with water (to be solved), and one with the solid porous matrix (not to be solved); (b) slowly-varying liquid phase fraction (\viz porosity); (c) no interphase mass transfer; and (d) interphase momentum transfer equating Darcy and Forchheimer forces.

This work clarifies the range of applicability of REV models—and by extension, the corresponding LBM implementations—situates REV modelling within the broader context of EE models, and deepens our understanding of hydraulic dispersion and its impact on REV modelling.
\section*{Author contribution statement}
\textbf{D.~Dapelo}: 
Conceptualization, 
Methodology, 
Formal analysis, 
Investigation, 
Writing - Original draft, 
Writing - Review {\&} Editing, 
\textbf{S.~Simonis}: 
Formal analysis,
Investigation,
Writing - Review {\&} Editing, 
\textbf{M.~Mousavi~Nezhad}: 
Formal analysis,
Investigation,
Writing - Review {\&} Editing, 
\textbf{M.~J.~Krause}: 
Writing - Review {\&} Editing, 
Supervision, 
Funding acquisition;
\textbf{J,~Bridgeman}: 
Writing - Review {\&} Editing, 
Supervision.
All authors read and approved the final manuscript.
\section*{Declaration of Generative AI and AI-assisted technologies in the writing process}
During the preparation of this work, the authors used ChatGPT for grammar checks and text polishing. After using this tool/service, the authors reviewed and edited the content as needed and take full responsibility for the content of the publication.

\bibliographystyle{abbrvnat}
\bibliography{library/main}

\end{document}